# General Exact Solution of Einstein Field Equations for Diagonal, Vacuum, Separable Metrics

Ron Lenk[1]

*Marietta, GA, USA*

September 28, 2010

**Abstract**

In this article we find the general, exact solution for the gravitational field equations for diagonal, vacuum, separable metrics. These are metrics each of whose terms can be separated into functions of each space-time variable separately. Other than this, the functions are completely arbitrary; no symmetries are assumed; no limitations are placed on the coordinates. There are 16 functions, which with specific selection of coordinates reduce to 12. Since there are 10 field equations, two functions in the solution are completely arbitrary. The field equations are solved exactly. The solution for each function is presented analytically, with a total of three parameters and ten constants in addition to the two arbitrary functions.

## 1 Introduction

Exact solutions of General Relativity are hard to come by. Almost all of those known depend on assuming symmetries of the space-time. Even these are not very general solutions, depending on a few independently variable continuous parameters such as mass, charge and angular momentum[2].

Since the field equations of General Relativity are partial differential equations (PDEs), general solutions should depend on independently variable functions. However, the non-linearity of the field equations makes them extremely hard to solve, and this has prevented much progress in this direction.

---

[1] e-mail: ron.lenk@reliabulb.com
[2] The situation is well summarized by Stephani et al. [1], "…we must first make some assumptions about $T_{ab}$. Beyond this we may proceed, for example, by imposing symmetry conditions on the metric, by restricting the algebraic structure of the Riemann tensor, by adding field equations for the matter variables or by imposing initial and boundary conditions. The exact solutions known have all been obtained by making some such restrictions."

In this paper we take a fresh look at this problem. Observing that many of the PDEs of physics can be solved using separation of variables, we look at a diagonal vacuum metric, and assume that the metric functions are separable. In a four-dimensional diagonal metric, there are four functions of the four coordinates. Assuming that each of these is separable yields a total of 4 x 4 = 16 functions, each function dependent on one and only one coordinate. We can impose four coordinate conditions on these due to general covariance, leaving us with $16 - 4 = 12$ functions. Finally, the Einstein tensor for this metric consists of ten independent expressions, all of which are non-zero. We use the vacuum field equations to set the tensor equal to zero, leaving $12 - 10 = 2$ functions. These remaining two functions are arbitrary. All of the other functions in the metric are determined by solving the field equations. No space-time symmetries are assumed, and there are no restrictions placed on the coordinates.

The fact that we can solve these equations is greatly aided by the fact that the field equations for a diagonal, separable variable metric don't contain mixed derivatives. We begin the solution with the observation that the off-diagonal terms of the Einstein tensor are first-order order PDEs. Each of them can be attacked by separation of variables, yielding sets of first-order ODEs. By carefully selecting which order to solve these in, we can simplify them by applying the solutions from the previous ones, easing the task. We first solve the off-diagonal time components, and then the off-diagonal space components.

With all of the off-diagonal components satisfied, we use the solutions to simplify the diagonal equations. These then turn out to be separable as well. The result is coupled sets of ODEs for the functions of each space-time coordinate. Again using judicious selection, and with the help of MAPLE [2], we simultaneously solve the sets. The solutions imply relationships between the separation variables, which aids in solving the next set. We solve all of the diagonal components, yielding the complete solution.

It should be noted that the solution we derive here is 'general' not 'complete'. In the various equations we solve, there are numerous separation constants. We only consider the cases in which the separation constants are generic, for example non-zero. In this sense the solution is 'general'. Looking instead at setting one or more of the separation constants to zero or other non-generic values results in hundreds of special solution cases, some of whose structure differs significantly from the general solution.

It might be wondered what is the use of this exercise? You can put any number of completely arbitrary functions into the metric, calculate the Einstein tensor, and this will be a 'solution' of the field equations by definition. However, this 'solution' is not useful, because such matter probably doesn't exist. This is not the case here. We insist that the matter content of the space-time is zero. This vacuum assumption very strongly constrains the solution. The solution presented here may be viewed as a small step towards a general solution of the vacuum field equations. As for usefulness, it could also, for example, be a useful test bed for numerical codes, since it has no symmetries whatsoever.

## 2 The Problem

The most general diagonal metric is given by

$$\begin{pmatrix} A(t,r,\theta,\phi) & 0 & 0 & 0 \\ 0 & B(t,r,\theta,\phi) & 0 & 0 \\ 0 & 0 & C(t,r,\theta,\phi) & 0 \\ 0 & 0 & 0 & D(t,r,\theta,\phi) \end{pmatrix} \quad (1)$$

with the coordinate system being [t, r, $\theta$, $\phi$]. Here we have assumed that the coordinate names are the same as those used in the case of spherical symmetry. However, there is no assumption of any symmetry, nor of compactification; we just find it convenient to use these coordinate names for convenience in knowing their position in the metric. In particular, we do not assume that either $\theta$ or $\phi$ is periodic, nor that r is strictly positive.

Next we add in the assumption of separability of the metric terms in Eqn. (1). This means that each function is a product of functions, each of which is a function of only one space-time coordinate. Re-using the letters A, B, C and D, the most general diagonal, separable variable metric is given by

$$\begin{pmatrix} A(t)B(r)C(\theta)D(\phi) & 0 & 0 & 0 \\ 0 & E(t)F(r)G(\theta)H(\phi) & 0 & 0 \\ 0 & 0 & J(t)K(r)L(\theta)M(\phi) & 0 \\ 0 & 0 & 0 & N(t)P(r)Q(\theta)S(\phi) \end{pmatrix} \quad (2)$$

where we have skipped I, O and R for reasons of typographical convenience.

Now general covariance allows us to select new coordinates t', r', $\theta$ and $\phi$'. We will make a selection of them to eliminate four of the functions in the metric Eqn. (2). We will make the following transform from [t, r, $\theta$, $\phi$] to [t', r', $\theta$', $\phi$']:

$$A(t)B(r)C(\theta)D(\varphi)dt^2 \to B'(r')dt'^2$$

$$E(t)F(r)G(\theta)H(\phi)dr^2 \to E'(t')F'(r')G'(\theta')H'(\phi')dr'^2$$

$$J(t)K(r)L(\theta)M(\varphi)d\theta^2 \to J'(t')r'^2L'(\theta')M'(\varphi')d\theta'^2 \qquad (3)$$

$$N(t)P(r)Q(\theta)S(\phi)d\varphi^2 \to N'(t')P'(r')Q'(\theta')S'(\phi')d\varphi'^2$$

Removing the primes from the new coordinates, we are left with the metric tensor:

$$g_{\mu\nu} = \begin{pmatrix} B(r) & 0 & 0 & 0 \\ 0 & E(t)F(r)G(\theta)H(\phi) & 0 & 0 \\ 0 & 0 & J(t)r^2L(\theta)M(\phi) & 0 \\ 0 & 0 & 0 & N(t)P(r)Q(\theta)S(\phi) \end{pmatrix} \qquad (4)$$

## 3 The Off-Diagonal Time Components

We begin the evaluation of the Einstein tensor with the metric Eqn. (4). The three off-diagonal time components are

$$G_{tr} = NP\left(-2EB\frac{dJ}{dt} + rE\frac{dJ}{dt}\frac{dB}{dr} + 2JB\frac{dE}{dt}\right) + rJ\left(-BE\frac{dN}{dt}\frac{dP}{dr} + PE\frac{dN}{dt}\frac{dB}{dr} + NB\frac{dE}{dt}\frac{dP}{dr}\right)$$

$$G_{t\theta} = NQ\frac{dG}{d\theta}\left(J\frac{dE}{dt} - E\frac{dJ}{dt}\right) + EG\frac{dQ}{d\theta}\left(J\frac{dN}{dt} - N\frac{dJ}{dt}\right) \qquad (5)$$

$$G_{t\varphi} = JM\frac{dH}{d\varphi}\left(E\frac{dN}{dt} - N\frac{dE}{dt}\right) + EH\frac{dM}{d\varphi}\left(J\frac{dN}{dt} - N\frac{dJ}{dt}\right)$$

where we have eliminated common pre-factors because we use the field equations to set each component equal to zero. Now we can do separation of variables. Starting with $G_{t\theta} = 0$,

$$NQ\frac{dG}{d\theta}\left(J\frac{dE}{dt}-E\frac{dJ}{dt}\right)=-EG\frac{dQ}{d\theta}\left(J\frac{dN}{dt}-N\frac{dJ}{dt}\right) \tag{6a}$$

$$\frac{N\left(J\dfrac{dE}{dt}-E\dfrac{dJ}{dt}\right)}{E\left(J\dfrac{dN}{dt}-N\dfrac{dJ}{dt}\right)}=\frac{-G\dfrac{dQ}{d\theta}}{Q\dfrac{dG}{d\theta}}=\lambda \tag{6b}$$

where the last step, as usual, is because a function only of time can equal a function only of $\theta$ only if both are equal to a constant, in this case $\lambda$. Solving the second equation in Eqn. (6b) first, we have

$$\frac{-G}{\left(\dfrac{dG}{d\theta}\right)}=\lambda\frac{Q}{\left(\dfrac{dQ}{d\theta}\right)} \tag{7}$$

and integrating this yields

$$G(\theta)=c_1 Q(\theta)^{-\frac{1}{\lambda}};\lambda\neq 0 \tag{8}$$

with $c_1$ an integration constant. $c_1$ cannot be zero, as we do not allow any of the metric components to be zero; this assumption will be used throughout, usually without explicit comment. As described in the introduction, we are ignoring the special case where $\lambda = 0$ in Eqn. (6b).

Since the first equation in Eqn. (6b) involves three functions, we for the moment attack the $R_{t\phi} = 0$ equation instead. Following the same reasoning, we have

$$JM\frac{dH}{d\varphi}\left(E\frac{dN}{dt}-N\frac{dE}{dt}\right)=-EH\frac{dM}{d\varphi}\left(J\frac{dN}{dt}-N\frac{dJ}{dt}\right) \tag{9a}$$

$$\frac{J\left(E\dfrac{dN}{dt}-N\dfrac{dE}{dt}\right)}{E\left(J\dfrac{dN}{dt}-N\dfrac{dJ}{dt}\right)}=\frac{-H\dfrac{dM}{d\varphi}}{M\dfrac{dH}{d\varphi}}=\mu \tag{9b}$$

and then

$$-\frac{\frac{dM}{d\varphi}}{M} = \mu \frac{\frac{dH}{d\varphi}}{H} \tag{10}$$

yielding

$$M(\varphi) = c_2 H(\varphi)^{-\mu} \tag{11}$$

We can now use the first parts of Eqns. (6b) and (9b) to solve for N(t) as a function of J(t). We have the two simultaneous equations

$$N\left(J\frac{dE}{dt} - E\frac{dJ}{dt}\right) = \lambda E\left(J\frac{dN}{dt} - N\frac{dJ}{dt}\right)$$

$$J\left(E\frac{dN}{dt} - N\frac{dE}{dt}\right) = \mu E\left(J\frac{dN}{dt} - N\frac{dJ}{dt}\right) \tag{12}$$

We isolate the terms in E separately from the terms in the derivative of E,

$$NJ\frac{dE}{dt} = E\left(\lambda J\frac{dN}{dt} - \lambda N\frac{dJ}{dt} + N\frac{dJ}{dt}\right)$$

$$-JN\frac{dE}{dt} = E\left(\mu J\frac{dN}{dt} - \mu N\frac{dJ}{dt} - J\frac{dN}{dt}\right) \tag{13}$$

We thus have

$$\frac{1}{E}\frac{dE}{dt} = \frac{1}{NJ}\left(\lambda J\frac{dN}{dt} - \lambda N\frac{dJ}{dt} + N\frac{dJ}{dt}\right) \tag{14a}$$

$$\frac{1}{E}\frac{dE}{dt} = \frac{-1}{JN}\left(\mu J\frac{dN}{dt} - \mu N\frac{dJ}{dt} - J\frac{dN}{dt}\right) \tag{14b}$$

These two equations can each be exactly integrated. Starting with Eqn. (14a), we have

$$\frac{dE}{E} = \lambda \frac{dN}{N} - \lambda \frac{dJ}{J} + \frac{dJ}{J}$$

$$\ln(E) = \lambda \ln(N) - \lambda \ln(J) + \ln(J) + c \tag{15}$$

$$E(t) = c_3 N(t)^\lambda J(t)^{1-\lambda}$$

and similarly for Eqn. (14b)

$$\frac{dE}{E} = -\mu \frac{dN}{N} + \mu \frac{dJ}{J} + \frac{dN}{N} \tag{16a}$$

$$\ln(E) = -\mu \ln(N) + \mu \ln(J) + \ln(N) + c \tag{16b}$$

$$E(t) = c_4 J(t)^\mu N(t)^{1-\mu} \tag{16c}$$

Since these two have to be equal, we must have $c_3 = c_4$, and $\lambda = 1 - \mu$. The solution for E is thus

$$E(t) = c_4 J(t)^\mu N(t)^{1-\mu}; \lambda = 1-\mu \tag{17}$$

We will be using µ instead of λ throughout from now on. With the solutions Eqns. (8), (11) and (17), the second and third equations of Eqn. (5) are completely solved. Turning now to the first equation of Eqn. (5), we start by substituting in Eqn. (17) for E:

$$\begin{aligned}
G_{tr} &= E\left[-2NPB\frac{dJ}{dt} + rNP\frac{dJ}{dt}\frac{dB}{dr} - rJB\frac{dN}{dt}\frac{dP}{dr} + rJP\frac{dN}{dt}\frac{dB}{dr}\right] + \frac{dE}{dt}\left[2NPJB + rJNB\frac{dP}{dr}\right] \\
&= c_4 J^\mu N^{1-\mu}\left[-2NPB\frac{dJ}{dt} + rNP\frac{dJ}{dt}\frac{dB}{dr} - rJB\frac{dN}{dt}\frac{dP}{dr} + rJP\frac{dN}{dt}\frac{dB}{dr}\right] + \\
&\quad \left[c_4\mu J^{\mu-1} N^{1-\mu}\frac{dJ}{dt} + c_4\mu J^\mu (1-\mu) N^{-\mu}\frac{dN}{dt}\right]\left[2NPJB + rJNB\frac{dP}{dr}\right] \\
&= N\frac{dJ}{dt}\left[rP\frac{dB}{dr} + 2(\mu-1)PB + \mu rB\frac{dP}{dr}\right] + J\frac{dN}{dt}\left[rP\frac{dB}{dr} + \mu(1-\mu)2PB + (\mu-\mu^2-1)rB\frac{dP}{dr}\right]
\end{aligned} \tag{18}$$

This is immediately separable,

$$\frac{N\frac{dJ}{dt}}{J\frac{dN}{dt}} = -\frac{rP\frac{dB}{dr} + \mu(1-\mu)2PB + (\mu-\mu^2-1)rB\frac{dP}{dr}}{rP\frac{dB}{dr} + 2(\mu-1)PB + \mu rB\frac{dP}{dr}} = \kappa \tag{19}$$

since the lhs is only a function of t, and the rhs is only a function of r. We can now solve immediately for the first of these two equations,

$$N(t) = c_5 J(t)^{\frac{1}{\kappa}}; \kappa \neq 0 \tag{20}$$

Substituting Eqn. (20) back into Eqn. (5a), we have the equation

$$2(-\kappa + \mu\kappa + 1 - \mu)PB + (\kappa+1)rP\frac{dB}{dr} + (\kappa-1)\mu rB\frac{dP}{dr} = 0 \tag{21}$$

which can be solved exactly for P in terms of B:

$$P(r) = c_6 r^{\frac{2(1-\mu)}{\mu}} B(r)^{-\frac{\kappa+1}{\mu(\kappa-1)}}; \kappa \neq 1; \mu \neq 0 \tag{22}$$

These four equations, Eqns. (8), (11), (17) and (22), solve all of the off-diagonal time components of the Einstein tensor.

## 4 The Other Off-Diagonal Components

Next we look at the other three off-diagonal terms of the Einstein tensor,

$$G_{r\theta} = rPQ\frac{dB}{dr}\frac{dG}{d\theta} - rBG\frac{dP}{dr}\frac{dQ}{d\theta} + rBQ\frac{dP}{dr}\frac{dG}{d\theta} + 2BPG\frac{dQ}{d\theta} \tag{23a}$$

$$G_{r\varphi} = -rMP\frac{dB}{dr}\frac{dH}{d\varphi} + 2BHP\frac{dM}{d\varphi} - 2BMP\frac{dH}{d\varphi} - rBH\frac{dP}{dr}\frac{dM}{d\varphi} \tag{23b}$$

$$G_{\theta\varphi} = -MQ\frac{dG}{d\theta}\frac{dH}{d\varphi} + HQ\frac{dG}{d\theta}\frac{dM}{d\varphi} + GM\frac{dH}{d\varphi}\frac{dQ}{d\theta} \tag{23c}$$

where we have again eliminated pre-factors. Starting with Eqn. (23a), we can eliminate G by using Eqn. (8):

$$G_{r\theta} = rQ\left(P\frac{dB}{dr} + B\frac{dP}{dr}\right)\frac{dG}{d\theta} + B\frac{dQ}{d\theta}\left(2P - r\frac{dP}{dr}\right)G$$

$$= rQ\left(P\frac{dB}{dr} + B\frac{dP}{dr}\right)\left(\frac{1}{\mu-1}c_1 Q^{\frac{1}{\mu-1}-1}\frac{dQ}{d\theta}\right) + B\frac{dQ}{d\theta}\left(2P - r\frac{dP}{dr}\right)\left(c_1 Q^{\frac{1}{\mu-1}}\right) \quad (24)$$

$$= c_1 Q^{\frac{1}{\mu-1}}\frac{dQ}{d\theta}\left[\frac{1}{\mu-1}r\left(P\frac{dB}{dr} + B\frac{dP}{dr}\right) + B\left(2P - r\frac{dP}{dr}\right)\right] = 0$$

There are two solutions to this. In the first, the terms outside the braces are zero, in which case $Q(\theta) = c_7$. The other is to solve the term inside the braces for zero, for which we divide by rPB.

$$\frac{1}{\mu-1}r\left(P\frac{dB}{dr} + B\frac{dP}{dr}\right) + B\left(2P - r\frac{dP}{dr}\right) = 0 \quad (25a)$$

$$\frac{1}{\mu-1}\left(\frac{dB}{B} + \frac{dP}{P}\right) + \left(\frac{2}{r}dr - \frac{dP}{P}\right) = 0 \quad (25b)$$

$$\frac{1}{\mu-1}(\ln(B) + \ln(P)) + (2\ln(r) - \ln(P)) = c$$

$$B^{\frac{1}{\mu-1}} P^{\frac{1}{\mu-1}} r^2 P^{-1} = c \quad (25c)$$

so that we have two regular solutions and one singular limit:

$$Q(\theta) = c_7 \quad (26a)$$

$$P(r) = c_8 r^{\frac{2(\mu-1)}{\mu-2}} B(r)^{\frac{1}{\mu-2}}; \mu \neq 2 \quad (26b)$$

$$B(r) = c_8 r^{-2}; \mu = 2 \quad (26c)$$

Of these, both Eqns. (26a) and (26b) are general. We already have an expression for P in Eqn. (22), and simultaneously satisfying both it and Eqn. (26b) requires P and B both to be a power of r (too complicated to be worth showing here). Since this determines two metric functions and Eqn. (26a) determines only one, we retain only Eqn. (26a) going forward.

We can now turn to Eqn. (23b), which has the same structure as Eqn. (23a). We eliminate M by using Eqn. (11):

$$G_{r\varphi} = BH\left(2P - r\frac{dP}{dr}\right)\frac{dM}{d\varphi} + P\frac{dH}{d\varphi}\left(-r\frac{dB}{dr} - 2B\right)M$$

$$= BH\left(2P - r\frac{dP}{dr}\right)\left(-\mu c_2 H^{-\mu-1}\frac{dH}{d\varphi}\right) + P\frac{dH}{d\varphi}\left(-r\frac{dB}{dr} - 2B\right)\left(c_2 H^{-\mu}\right) \qquad (27)$$

$$= c_2 H^{-\mu}\frac{dH}{d\varphi}\left[-\mu B\left(2P - r\frac{dP}{dr}\right) + P\left(-r\frac{dB}{dr} - 2B\right)\right] = 0$$

We again have two solutions to this. In the first, the terms outside the braces are zero, in which case $H(\varphi) = c_{10}$. The other is to solve the term inside the braces for zero, for which we again divide by rBP:

$$-\mu\left(\frac{2}{r}dr - \frac{dP}{P}\right) + \left(-\frac{dB}{B} - \frac{2}{r}dr\right) = 0 \qquad (28)$$

$$2\mu\ln(r) - \mu\ln(P) + \ln(B) + 2\ln(r) = c$$

which gives rise to the following two general cases and one singular case:

$$H(\varphi) = c_{10} \qquad (29a)$$

$$P(r) = c_{11} r^{\frac{2(1+\mu)}{\mu}} B(r)^{\frac{1}{\mu}}; \mu \neq 0 \qquad (29b)$$

$$B(r) = c_{11} r^{-2}; \mu = 0 \qquad (29c)$$

Of these, Eqns. (29a) and (29b) are both general. We already have an expression for P in Eqn. (22), and simultaneously satisfying both it and Eqn. (29b) requires P and B both to be a power of r. Since this determines two metric functions and Eqn. (29a) determines only one, we retain only Eqn. (29a) going forward.

Finally, using both Eqns. (8) and (11) in Eqn. (23c), we have

$$G_{\theta\varphi} = -MQ\frac{dG}{d\theta}\frac{dH}{d\varphi} + HQ\frac{dG}{d\theta}\frac{dM}{d\varphi} + GM\frac{dH}{d\varphi}\frac{dQ}{d\theta}$$

$$= -c_2 H^{-\mu} Q\left(c_1 \frac{1}{\mu-1} Q^{\frac{1}{\mu-1}-1} \frac{dQ}{d\theta}\right)\frac{dH}{d\varphi} + HQ\left(c_1 \frac{1}{\mu-1} Q^{\frac{1}{\mu-1}-1} \frac{dQ}{d\theta}\right)\left(-c_2 \mu H^{-\mu-1}\frac{dH}{d\varphi}\right) \quad (30)$$

$$+ c_1 c_2 H^{-\mu} Q^{\frac{1}{\mu-1}} \frac{dH}{d\varphi}\frac{dQ}{d\theta} = -\frac{2c_1 c_2}{\mu-1} H^{-\mu} Q^{\frac{1}{\mu-1}} \frac{dH}{d\varphi}\frac{dQ}{d\theta} = 0$$

Thus, either H or Q or both must be constant. Both of these are already true because of Eqns. (26a) and (29a).

This then completes the solution of all of the off-diagonal equations. We collect all of the parts here.

$$G(\theta) = c_1 Q(\theta)^{-\frac{1}{\lambda}} \tag{8}$$

$$M(\varphi) = c_2 H(\varphi)^{-\mu} \tag{11}$$

$$E(t) = c_4 J(t)^{\mu} N(t)^{1-\mu} \tag{17}$$

$$N(t) = c_5 J(t)^{\frac{1}{\kappa}}; \kappa \neq 0 \tag{20}$$

$$P(r) = c_6 r^{\frac{2(1-\mu)}{\mu}} B(r)^{-\frac{\kappa+1}{\mu(\kappa-1)}}; \kappa \neq 1; \mu \neq 0 \tag{22}$$

$$Q(\theta) = c_7 \tag{26a}$$

$$H(\varphi) = c_{10} \tag{29a}$$

$$H(\varphi) = c_{10} \text{ or } Q(\theta) = c_7 \tag{30}$$

Simplifying these, we end up with the solution for the off-diagonal components:

$$E(t) = c_4 J(t)^{\mu + \frac{1-\mu}{\kappa}} \tag{31a}$$

$$G(\theta) = c_1 \tag{31b}$$

$$H(\varphi) = c_{10} \tag{31c}$$

$$M(\varphi) = c_2 \tag{31d}$$

$$N(t) = c_5 J(t)^{\frac{1}{\kappa}}; \kappa \neq 0 \tag{31e}$$

$$P(r) = c_6 r^{\frac{2(1-\mu)}{\mu}} B(r)^{\frac{\kappa+1}{\mu(\kappa-1)}}; \kappa \neq 1; \mu \neq 0 \tag{31f}$$

$$Q(\theta) = c_7 \tag{31g}$$

To avoid the distinct possibility of having introduced errors, we have directly substituted this partial solution into the field equations (using Maple), leaving the other functions in the metric Eqn. (4) arbitrary, and verified that all of the off-diagonal terms of the Einstein tensor are identically zero.

## 5 The Diagonal Components

In this section we use the partial solution Eqn. (31) to replace E, G, H, M, N, P and Q in the diagonal components of the Einstein tensor. The diagonal components with these substitutions are:

$$G_{tt} = 0 = c_1 c_4 c_{10} r^2 BF^2 J^{\frac{\kappa\mu - \mu + 1}{\kappa}} \mu^2 \left(\kappa^4 \mu - 3\kappa^2 + 1 + 2\mu\kappa - 2\mu\kappa^3 + 2\kappa^3 - \mu\right)\left(\frac{dJ}{dt}\right)^2$$
$$+ 4J^2 FB^2 \kappa^2 \left(4\mu\kappa - 2\kappa^2\mu - 2\kappa\mu^2 + \kappa^2\mu^2 - 2\mu + \mu^2 - 2\kappa + \kappa^2 + 1\right)$$
$$- rJ^2 \kappa^2 \left[\begin{array}{l} 2\mu(-2\kappa + \kappa^2 + 1)B^2 \dfrac{dF}{dr} + \mu(-\kappa^2 + 1)rB \dfrac{dF}{dr}\dfrac{dB}{dr} + 2\mu(\kappa^2 - 1)rFB\dfrac{d^2 B}{dr^2} \\ + 2(\mu + 2\kappa^2 - 2 - \kappa^2\mu)BF\dfrac{dB}{dr} + (2\mu - 1 - 2\kappa^2\mu - \kappa^2 - 2\kappa)rF\left(\dfrac{dB}{dr}\right)^2 \end{array}\right] \tag{32}$$

$$G_{rr} = 0 = -c_1 c_4 c_{10} \mu r^2 BFJ^{\frac{\kappa\mu-\mu+1}{\kappa}}\left[\left(-2\kappa^3+2\kappa\right)J\frac{d^2J}{dt^2}+\left(\kappa^3-2\kappa+1\right)\left(\frac{dJ}{dt}\right)^2\right]$$
$$+4B^2J^2\kappa^2\left(-\kappa\mu+\kappa-1+\mu\right)-\kappa^2 rJ^2\left(4B\frac{dB}{dr}+(\kappa+1)r\left(\frac{dB}{dr}\right)^2\right)$$
(33)

$$G_{\theta\theta} = 0 = c_1 c_4 c_{10} r^2 BF^2 J^{\frac{\mu\kappa-\mu+1}{\kappa}}\mu^2\left[\begin{array}{l}2\kappa\left(-3\kappa^2\mu+3\kappa\mu-4\kappa-\mu+2+2\kappa^2+\kappa^3\mu\right)J\frac{d^2J}{dt^2}\\+\left(\begin{array}{l}-15\kappa^2\mu+3-3\mu-4\kappa^3+9\kappa^3\mu+\mu^2\\+\mu^2\kappa^4+6\mu^2\kappa^2+11\kappa^2-10\kappa-4\mu^2\kappa^3\\-4\mu^2\kappa+11\kappa\mu-2\mu\kappa^4\end{array}\right)\left(\frac{dJ}{dt}\right)^2\end{array}\right]$$
$$+4\kappa^2 B^2 FJ^2\left(-4\kappa\mu^2+2\mu^2+2\kappa^2\mu^2+6\kappa\mu-3\kappa^2\mu-3\mu+\kappa^2-2\kappa+1\right)$$
$$+\kappa^2 rJ^2\left[\begin{array}{l}\left(\begin{array}{l}-\mu+\kappa^2\mu-\mu^2\kappa^2\\+2\mu^2\kappa-\mu^2\end{array}\right)rB\frac{dF}{dr}\frac{dB}{dr}+2\left(\begin{array}{l}-2\kappa^2-\mu-\mu^2+3\kappa^2\mu\\-2\kappa\mu-\mu^2\kappa^2+2\mu^2\kappa+2\end{array}\right)BF\frac{dB}{dr}\\+2\mu\left(\begin{array}{l}-2\mu\kappa+\mu\kappa^2\\-\kappa^2+2\kappa-1+\mu\end{array}\right)B^2\frac{dF}{dr}+2\mu\left(\begin{array}{l}1-\kappa^2+\mu\\-2\mu\kappa+\mu\kappa^2\end{array}\right)rBF\frac{d^2B}{dr^2}\\+\left(-\mu^2-\mu+\kappa^2+2\kappa-\mu^2\kappa^2+2\mu^2\kappa+\mu\kappa^2+1\right)rF\left(\frac{dB}{dr}\right)^2\end{array}\right]$$
(34)

$$G_{\varphi\varphi} = 0 = c_1 c_4 c_{10} r^{\frac{2-\mu}{\mu}} B^{\frac{\kappa\mu-\mu-\kappa-1}{\mu(\kappa-1)}} F^2 J^{\frac{1}{\kappa}}\left[\begin{array}{l}2\kappa(\kappa+\kappa\mu-\mu+1)J\frac{d^2J}{dt^2}+\\\left(\begin{array}{l}-\kappa^2+\mu^2\kappa^2-\mu\kappa^2-2\mu^2\kappa\\+3\kappa\mu-\kappa-2\mu+1+\mu^2\end{array}\right)\left(\frac{dJ}{dt}\right)^2\end{array}\right]$$
$$-r^{\frac{2-2\mu}{\mu}} B^{\frac{\kappa\mu-\mu-\kappa-1}{\mu(\kappa-1)}-1} J^{\frac{2\kappa-\kappa\mu+\mu}{\kappa}}\kappa^2\left(-2BF\frac{dB}{dr}+rF\left(\frac{dB}{dr}\right)^2+rB\frac{dB}{dr}\frac{dF}{dr}+2B^2\frac{dF}{dr}-2rBF\frac{d^2B}{dr^2}\right)$$
(35)

Eqn. (32) is immediately separable:

$$c_1 c_4 c_{10} J^{\frac{\kappa\mu-\mu+1}{\kappa}-2}\mu^2\left(\kappa^4\mu-3\kappa^2+1+2\mu\kappa-2\mu\kappa^3+2\kappa^3-\mu\right)\left(\frac{dJ}{dt}\right)^2 =$$
$$-\frac{4B\kappa^2}{r^2 F}\left(4\mu\kappa-2\kappa^2\mu-2\kappa\mu^2+\kappa^2\mu^2-2\mu+\mu^2-2\kappa+\kappa^2+1\right)$$
$$+\frac{\kappa^2}{rBF^2}\left[\begin{array}{l}2\mu\left(-2\kappa+\kappa^2+1\right)B^2\frac{dF}{dr}+\mu\left(-\kappa^2+1\right)rB\frac{dF}{dr}\frac{dB}{dr}+2\mu\left(\kappa^2-1\right)rFB\frac{d^2B}{dr^2}\\+2\left(\mu+2\kappa^2-2-\kappa^2\mu\right)BF\frac{dB}{dr}+\left(2\mu-1-2\kappa^2\mu-\kappa^2-2\kappa\right)rF\left(\frac{dB}{dr}\right)^2\end{array}\right]=\rho$$
(36)

The equation for J can be solved at once,

$$J(t) = \left[ \frac{\rho(\mu\kappa - \mu + 1)^2 (t - c_{11})^2}{4\kappa^2 c_1 c_4 c_{10} \mu^2 (\kappa - 1)^2 (\mu\kappa^2 + 2\kappa + 1 - \mu)} \right]^{\frac{\kappa}{\mu\kappa - \mu + 1}} \quad (37)$$

Eqn. (33) is also separable,

$$c_1 c_4 c_{10} \mu J^{\frac{\kappa\mu - \mu + 1}{\kappa} - 2} \left[ 2\kappa(1 - \kappa^2) J \frac{d^2 J}{dt^2} + (\kappa^3 - 2\kappa + 1) \left( \frac{dJ}{dt} \right)^2 \right] =$$
$$\frac{4B\kappa^2}{r^2 F}(-\kappa\mu + \kappa - 1 + \mu) - \frac{\kappa^2}{rBF} \left( 4B \frac{dB}{dr} + (\kappa + 1) r \left( \frac{dB}{dr} \right)^2 \right) = \sigma \quad (38)$$

We substitute solution Eqn. (37) into the J equation of Eqn. (38), and find that it is satisfied if

$$\sigma = \frac{\rho(\mu\kappa^2 - \kappa^2 - \mu)}{\mu(\mu\kappa^3 + 2\kappa^2 - \mu\kappa^2 - \kappa - \kappa\mu + \mu - 1)} \quad (39)$$

Continuing on, Eqn. (34) is also separable

$$c_1 c_4 c_{10} J^{\frac{\mu\kappa - \mu + 1}{\kappa} - 2} \mu^2 \left[ \begin{array}{l} 2\kappa(-3\kappa^2\mu + 3\kappa\mu - 4\kappa - \mu + 2 + 2\kappa^2 + \kappa^3\mu) J \dfrac{d^2 J}{dt^2} \\ + \left( \begin{array}{l} -15\kappa^2\mu + 3 - 3\mu - 4\kappa^3 + 9\kappa^3\mu + \mu^2 \\ + \mu^2\kappa^4 + 6\mu^2\kappa^2 + 11\kappa^2 - 10\kappa - 4\mu^2\kappa^3 \\ - 4\mu^2\kappa + 11\kappa\mu - 2\mu\kappa^4 \end{array} \right) \left( \dfrac{dJ}{dt} \right)^2 \end{array} \right] =$$
$$-\frac{4\kappa^2 B}{r^2 F}(-4\kappa\mu^2 + 2\mu^2 + 2\kappa^2\mu^2 + 6\kappa\mu - 3\kappa^2\mu - 3\mu + \kappa^2 - 2\kappa + 1)$$
$$-\frac{\kappa^2}{rBF^2} \left[ \begin{array}{l} \left( \begin{array}{l} -\mu + \kappa^2\mu - \mu^2\kappa^2 \\ + 2\mu^2\kappa - \mu^2 \end{array} \right) rB \dfrac{dF}{dr} \dfrac{dB}{dr} + 2 \left( \begin{array}{l} -2\kappa^2 - \mu - \mu^2 + 3\kappa^2\mu \\ -2\kappa\mu - \mu^2\kappa^2 + 2\mu^2\kappa + 2 \end{array} \right) BF \dfrac{dB}{dr} \\ + 2\mu \left( \begin{array}{l} -2\mu\kappa + \mu\kappa^2 \\ -\kappa^2 + 2\kappa - 1 + \mu \end{array} \right) B^2 \dfrac{dF}{dr} + 2\mu \left( \begin{array}{l} 1 - \kappa^2 + \mu \\ -2\mu\kappa + \mu\kappa^2 \end{array} \right) rBF \dfrac{d^2 B}{dr^2} \\ + (-\mu^2 - \mu + \kappa^2 + 2\kappa - \mu^2\kappa^2 + 2\mu^2\kappa + \mu\kappa^2 + 1) rF \left( \dfrac{dB}{dr} \right)^2 \end{array} \right] = \varsigma \quad (40)$$

We substitute solution Eqn. (37) into the J equation of Eqn. (40), and find that it is satisfied if

$$\varsigma = \frac{\rho}{\mu\kappa^2 + 2\kappa - \mu + 1} \qquad (41)$$

Finally, we separate Eqn. (35)

$$c_1 c_4 c_{10} J^{\frac{1-2\kappa+\kappa\mu-\mu}{\kappa}} \left[ 2\kappa(\kappa + \kappa\mu - \mu + 1)J\frac{d^2 J}{dt^2} + \begin{pmatrix} -\kappa^2 + \mu^2\kappa^2 - \mu\kappa^2 - 2\mu^2\kappa \\ +3\kappa\mu - \kappa - 2\mu + 1 + \mu^2 \end{pmatrix}\left(\frac{dJ}{dt}\right)^2 \right] =$$

$$\frac{\kappa^2}{rF^2}\left(-2BF\frac{dB}{dr} + rF\left(\frac{dB}{dr}\right)^2 + rB\frac{dB}{dr}\frac{dF}{dr} + 2B^2\frac{dF}{dr} - 2rBF\frac{d^2 B}{dr^2}\right) = \tau \qquad (42)$$

We substitute solution Eqn. (37) into the J equation of Eqn. (42), and find that it is satisfied if

$$\tau = \frac{\rho\kappa^2}{\mu^2(\mu\kappa^2 + 2\kappa - \mu + 1)(\kappa - 1)^2} \qquad (43)$$

With the values for the separation constants in terms of ρ in Eqns. (39), (41) and (43), the solution Eqn. (37) for J satisfies all of the diagonal equations.

The final step is to solve the r equations. In this case, we substitute the value of ρ from Eqn. (36) into Eqns. (38), (40) and (42), using the values of σ, ζ and τ from Eqns. (39), (41) and (43). These equations can then be simultaneously solved (using Maple and after a great deal of simplification) as

$$F(r) = c_{12} r^{-2\frac{\mu\kappa^2 - \mu + 2\kappa}{\kappa(\kappa+1)}} \qquad (44a)$$

$$B(r) = -\frac{c_{12}\rho}{4\mu^2(\kappa-1)^2(\mu\kappa^2 - \mu + 2\kappa + 1)} r^{-2\frac{\mu\kappa^2 - \mu + 2\kappa}{\kappa(\kappa+1)} + 2} \qquad (44b)$$

We have now solved all of the field equations. We combine Eqns. (31), (37) and (44) to end up with the complete solution (some of the constants have been renumbered):

$$B(r) = -\frac{c_3 \rho}{4\mu^2(\kappa-1)^2(\mu\kappa^2-\mu+2\kappa+1)} r^{-2\frac{\mu\kappa^2-\mu+2\kappa}{\kappa(\kappa+1)}+2} \tag{45a}$$

$$E(t) = \frac{\rho(\mu\kappa-\mu+1)^2(t-c_5)^2}{4\kappa^2 c_1 c_6 \mu^2(\kappa-1)^2(\mu\kappa^2+2\kappa+1-\mu)} \tag{45b}$$

$$F(r) = c_3 r^{-2\frac{\mu\kappa^2-\mu+2\kappa}{\kappa(\kappa+1)}} \tag{45c}$$

$$G(\theta) = c_1 \tag{45d}$$

$$H(\varphi) = c_{10} \tag{45e}$$

$$J(t) = \left[\frac{\rho(\mu\kappa-\mu+1)^2(t-c_5)^2}{4\kappa^2 c_1 c_4 c_6 \mu^2(\kappa-1)^2(\mu\kappa^2+2\kappa+1-\mu)}\right]^{\frac{\kappa}{\mu\kappa-\mu+1}} \tag{45f}$$

$$M(\varphi) = c_2 \tag{45g}$$

$$N(t) = c_8 \left[\frac{\rho(\mu\kappa-\mu+1)^2(t-c_5)^2}{4\kappa^2 c_1 c_4 c_6 \mu^2(\kappa-1)^2(\mu\kappa^2+2\kappa+1-\mu)}\right]^{\frac{1}{\mu\kappa-\mu+1}} \tag{45h}$$

$$P(r) = c_9 r^{\frac{2(1-\mu)}{\mu}} B(r)^{\frac{\kappa+1}{\mu(\kappa-1)}} \tag{45i}$$

$$Q(\theta) = c_7 \tag{45j}$$

$$\kappa \neq 0, 1, -1; \mu \neq 0; \rho \neq 0$$

As anticipated, the solution contains two arbitrary functions, L(θ) and S(ϕ). It also has three parameters κ, ρ and u, plus 10 constants $c_1 - c_{10}$. We have directly substituted this solution into the field equations (using Maple), leaving L and S arbitrary, and verified that all of the terms of the Einstein tensor are identically zero and confirming that no errors have been accidentally introduced.

## 7 Discussion

While a comprehensive study of this solution would be very interesting, it must be deferred. We make a few initial comments. All of the r-dependent functions (B, F and P) scale as monomials in r. Through suitable selection of μ and κ, a huge variety of integer and fractional exponents can be chosen. There is thus nothing special about the exponents of the functions in the metric in the Reissner-Nordstrom with cosmological constant solution, $r^{-1}$, $r^{-2}$ and $r^2$. They just happened to be easier to find.

We observe that this metric cannot be reduced to the form of the Schwarzschild solution, nor to that of flat space-time. Because of the high degree of symmetry of these solutions, this was to be expected. They correspond to some of the 'special' cases that were not explicitly presented here.

The t-dependent functions (E, J and N) are power-law in time. They cannot be made into constants with any acceptable set of parameters. Again, metrics without time dependence correspond to 'special' cases. The actual behavior of the metric functions is either expanding or contracting, depending on the specific values of μ and κ.

We have not assumed any limitations on the coordinates. Compactification of coordinates could be achieved, for example, by selecting L or S to be sine squared, or any other periodic function. There are apparent singularities at r = 0 and at t = $c_5$. These are coordinate, not real, singularities, as the Kretschmann tensor for this metric is zero. Thus, the r coordinate can have the range -∞ to +∞. Of course, it could also be limited to non-negative values.

There are a number of directions for future work. The metric shows some resemblance to the Kasner metric, and it should be determined if it reduces to it with a coordinate transformation and in a suitable limit. The special cases that were skipped here should be pursued, finding in particular those that correspond to Minkowski and Schwarzschild space-times. A classification of the special solutions would also be interesting.

## Acknowledgements

I would like to thank Pablo Laguna for a discussion.

## References

[1] H. Stephani, D. Kramer, M. MacCallum, C. Hoenselaers, E. Herlt, Exact Solutions of Einstein's Field Equations, 2nd Edition, Cambridge University Press (2003).

[2] A product of Waterloo Maple Inc., Waterloo, Ontario, Canada (see http://www.maplesoft.com).